\documentclass[english,aps,prb,reprint, superscriptaddress, amsmath]{revtex4-2}

\usepackage{graphicx}% Include figure files
\usepackage{dcolumn}% Align table columns on decimal point
\usepackage{bm}% bold math
\usepackage[colorlinks,bookmarks=false,citecolor=blue,linkcolor=blue,urlcolor=blue]{hyperref}
\usepackage{ulem}
\usepackage{xcolor}

\begin{document}

%\preprint{APS/123-QED}

\title{Dynamics of Gate-Controlled Superconducting Dayem Bridges}% Force line breaks with \\

\author{François Joint}
\email{joint@chalmers.se}
\affiliation{
	Department of Microtechnology and Nanoscience, Chalmers University of Technology, 412 96 Gothenburg, Sweden
}%
\author{Kazi Rafsanjani Amin}%
\affiliation{
	Department of Microtechnology and Nanoscience, Chalmers University of Technology, 412 96 Gothenburg, Sweden
}%
\author{Ivo Cools}
\affiliation{
	Department of Microtechnology and Nanoscience, Chalmers University of Technology, 412 96 Gothenburg, Sweden
}%
\author{Simone Gasparinetti}
\affiliation{
	Department of Microtechnology and Nanoscience, Chalmers University of Technology, 412 96 Gothenburg, Sweden
}%

\begin{abstract}
Local control of superconducting circuits by high-impedance electrical gates offers potential advantages in superconducting logic, quantum processing units, and cryoelectronics. Recent experiments have reported gate-controlled supercurrent in Dayem bridges made of metallic superconductors, mediated by direct current leakage, out-of-equilibrium phonons, or possibly other mechanisms. However, a time-domain characterization of this effect has been lacking. Here, we integrate Dayem bridges made of Niobium on Silicon into coplanar-waveguide resonators, and measure the effect of the gate voltage at steady state and during pulsed operation. We consider two types of arrangements for the gate: a side-coupled gate
and a remote injector. In both cases, we observe sizable changes in the real and the imaginary part of the constriction's impedance for gate voltages of the order of 1~V. However, we find striking differences in the time-domain dynamics, with the remote injector providing a faster and more controlled response. Our results contribute to our understanding of gate-controlled superconducting devices and their suitability for applications.
\end{abstract}

\maketitle

\section{Introduction}
In recent years, the potential of voltage-controlled superconducting logic has garnered significant attention due to its promising integration with Complementary Metal-Oxide-Semiconductor (CMOS)\cite{acharya2023} technologies and potential superiority over rapid single flux quantum (RSFQ) logic. Numerous studies\cite{desimoni2018, elalaily2021,ritter2021,catto_microwave_2022,alegria_high-energy_2021,basset_gate-assisted_2021} have demonstrated reversible switching of a superconducting nanowire between superconducting and normal (metallic) state by applying a gate voltage ($V_G$). However, the microscopic mechanism responsible for this effect remains unclear and highly debated \cite{ritter2021, golokolenov2021, ritter2022}. Given the wide range of superconductors, substrates, fabrication processes, and characterization techniques employed across the studies, it has been suggested that various mechanisms may contribute to the effect~\cite{ruf_effects_2023}. 

Most studies of gate-controlled superconductivity focused on dc measurements of the switching current of the nanowire under varying gate voltage and magnetic field. In a few studies~\cite{golokolenov2021,catto_microwave_2022,ryu_utilizing_2024}, the complex impedance of the nanowire at microwave frequencies has been probed by embedding the nanowire in a coplanar waveguide resonator coupled to a feedline. In these studies, it was found that for increasing gate voltage, the kinetic inductance of the nanowires increases, resulting in a decrease in the resonator's frequency, and, at the same time, dissipation in the circuit also increases, resulting in a decrease of the resonator's quality factor.

Notwithstanding these studies, little is known about the response of gate-tunable superconducting nanowires to a time-dependent gate voltage. In Ref.~\cite{ritter2021}, a switching time of 100~ns was reported for gated TiN nanowires on Si, but this figure was limited by the low-frequency setup used in the measurements. More comprehensive time-domain studies are thus needed. On one hand, time-domain studies provide additional information that may help elucidate the microscopic mechanisms beyond the effect. On the other hand, they can be used to assess the suitability of gate-tunable superconducting elements for applications such as superconducting logic or fast microwave switching. 

In this letter, we explore the dynamic properties of the field effect on a superconducting nanowire (Dayem) embedded in a $\lambda/4$ superconducting microwave resonator. We probe the system's dynamics by introducing a perturbation, specifically by applying a gate pulse to the superconducting Dayem, followed by the time-resolved measurement of the coherent scattering parameters of the resonators. Our study focuses on two distinct gate configurations and geometry: a direct quasiparticle injection gate (finger gate) and a phonon-mediated pair-breaking gate (remote gate). Our findings reveal distinct time scales of response, which we attribute to the different mechanisms activated by the gate configuration. The direct quasiparticle injection method induces hot spots within the nanowire, leading to a relatively slower switching process due to prolonged quasiparticle diffusion, scattering, and recombination times. Conversely, the phonon-mediated pair-breaking approach, facilitated by remote gate electrodes, displays a markedly faster response, underscoring its efficiency.

\begin{figure*}
\includegraphics{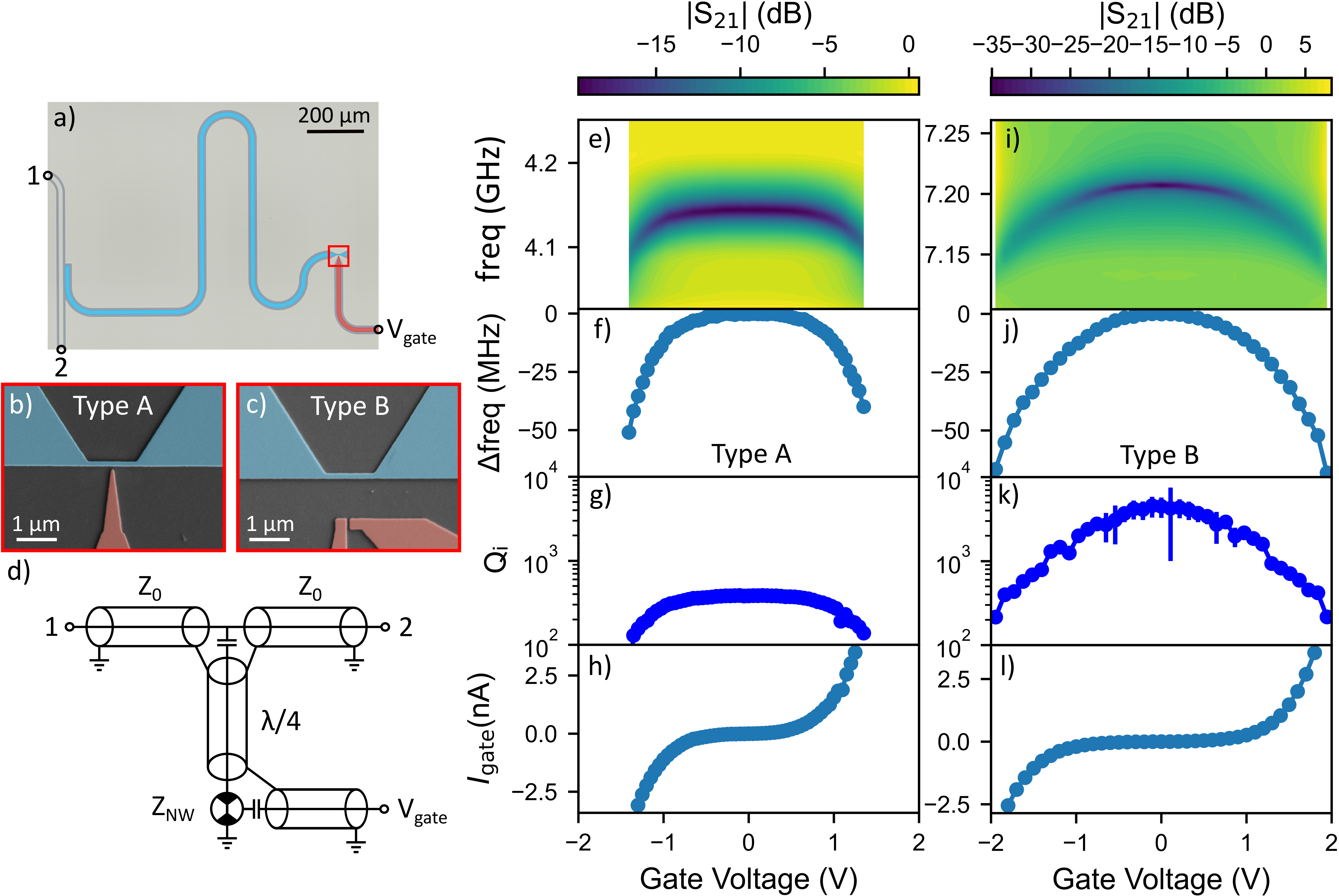}
\caption{\label{fig1} \textbf{Measured devices and CW measurements}. \textbf{a}, Optical image of one quarter-wave coplanar waveguide resonator capacitively coupled to a feedline (ports 1 and 2) and shunted to the ground through a $1.2\,\mu m$ long NW. The scanning electron microscope (SEM) image of the NW shunt is shown in \textbf{b} with a finger-type gate (Type A) and with a remote gate in \textbf{c}  (Type B). \textbf{d}, Circuit diagram of a measured device. The transmission line resonator with impedance $Z_0 = 50\Omega$ is shunted to ground by a NW with impedance $Z_{NW}$, which is gate controlled ($V_{gate}$), and capacitively coupled to the feedline with characteristic impedance $Z_0$. \textbf{e}, Amplitude of the scattering parameter $S_{21}$ in the function of the applied $V_{gate}$ for type A gate device. \textbf{f}, \textbf{g}, Internal quality factor and frequency shift as a function of $V_{gate}$ extracted from fits to $S_{21}$ in \textbf{e}. \textbf{h}, Gate current $I_{gate}$ as a function of $V_{gate}$ simultaneously to the data in \textbf{e}, \textbf{f} and \textbf{g}. \textbf{i}, Amplitude of $S_{21}$ for a device with type B gate. We extract frequency shift (\textbf{j}) and the internal quality factor (\textbf{k}) from \textbf{i}. \textbf{l}, Gate leakage current measured simultaneously to the data in \textbf{i}.}
\end{figure*}

\section{Experimental Set-up} 
The architecture of our devices consists of a superconducting thin film of Nb, 20nm in thickness, patterned into a quarter-wave coplanar waveguide (CPW) microwave resonator, and shunted to the ground with a $1.2~\mu\rm{m}$ long and 120~nm wide nanowire made of the same material [Fig.~\ref{fig1}(a-c)].  The center conductor of the CPW resonator is gradually tapered into the nanowire section to avoid any current crowding effects caused by abrupt width variations~\cite{clem_geometry-dependent_2011}. The resonators are capacitively coupled to a feedline and probed with a coherent microwave tone. Additionally, a gate electrode [colored red in Fig.~\ref{fig1}(a-c)] is connected via a 50~$\Omega$ impedance matched CPW waveguide, allowing the application of both static and pulsed voltage $V_G$ to the nanowire. Multiple resonators are connected to the same feedline using frequency-domain multiplexing.

This investigation employs two distinct gate geometries: a finger-type gate (referred to as type A [Fig.~\ref{fig1}(b)])~\cite{desimoni2018, elalaily2021, puglia2021, ritter2021, koch_gate-controlled_nodate},  and a remote-electrode gate (referred to as type B, [Fig.~\ref{fig1}(c)])~\cite{ritter2021, ritter2022, ryu_utilizing_2024}. The type A gate is positioned with a 70 nm gap from the nanowire, while the type B gate is spaced 70 nm apart from the ground plane and set at a distance of $1~\mu\rm{m}$  from the nanowire, leading to different electric field distribution around the gates. In type A gate, the electric field is confined between the gate electrode and the nanowire. In contrast in type B, the electric field is largely confined between the gate electrode and the adjacent ground electrode, with minimal extension to the nanowire~\cite{ritter2022}. As a gate voltage is applied, the inductance of the nanowire $L_{NW}$ is modified and, consequently, the resonator's resonant frequency $f_r$ is shifted \cite{splitthoff_gate-tunable_2022}. Strategically positioning the nanowire at the resonator's current anti-node amplifies the effect of $L^{NW}_K$ variation on $f_r$, while the placement at the voltage node minimizes coupling between the resonator and the gate electrode. All the measurements are conducted at the base temperature ($T_{\rm{mx}}\sim15$~mK) of a dilution refrigerator.

\section{Gate Tunabeable Resonator}
In Fig.~\ref{fig1}(e, f), we plot the magnitude of complex transmission parameter  $|S_{21}|$ versus gate voltage $V_{G}$, across $f_r$ for two devices with type-A and type-B gate respectively. By fitting the resonance traces~\cite{probst_efficient_2015} for each $V_G$, we 
extract the resonance frequencies ($f_r$) and internal quality factors ($Q_i$). We observe a monotonic decrease in both $f_r$ [Fig.~\ref{fig1}(g, j)]  and $Q_i$ [Fig.~\ref{fig1}(g, k)] with increasing $V_G$. We note an increase in the leakage current $I_{\rm{gate}}$ with increasing $V_G$, and we limit our measurements to $|I_{\rm{gate}}| \leq 2.5$~nA, thus $\lvert 1.2 \rvert V$ for type-A and $\lvert 2 \rvert V$ for type-B. This observed decrease of $f_r$ can be attributed to an enhancement in the kinetic inductance of the nanowire ($L^{NW}_K$) due to a decrease in the superfluid density, which contributes to an augmented total inductance of the resonator \cite{splitthoff_gate-tunable_2022}. Through electromagnetic (EM) simulations that model the nanowire as a lumped inductor while reflecting the precise device geometry of the distributed elements, we can predict the variation in $f_r$ corresponding to changes in the nanowire's inductance. By comparing the EM simulations with measurements, we obtain a $\Delta L^{NW}_K=16$~pH for $\max|\Delta f_{r}|=50$ ~MHz for type-A gated, and very similar $\Delta L^{NW}_K=14$~pH for $\max|\Delta f_{r}|=60$ ~MHz for type-B gated devices.

We observe a distinct difference in the trends of $V_G$ dependence on $f_r$ and $Q_i$ between the two device types. For type-A device, there is a gradual decline in $f_r$ with increasing $|V_G|$, which accelerates once $|V_G| >1$~V. A similar pattern is observed in $Q_i$ also. Conversely, the type-B device demonstrates a consistent shift in both $f_r$ and $Q_i$ across the full $|V_G|$, exhibiting a near-parabolic dependence on $|V_G|$.

Furthermore, a strong correlation between the observed trends in $f_r$ with the leakage current (Supplementary Fig.~S3) suggests that the leakage current plays an important role in the voltage tunability of superconductivity. 

\begin{figure*}[!ht]
\includegraphics{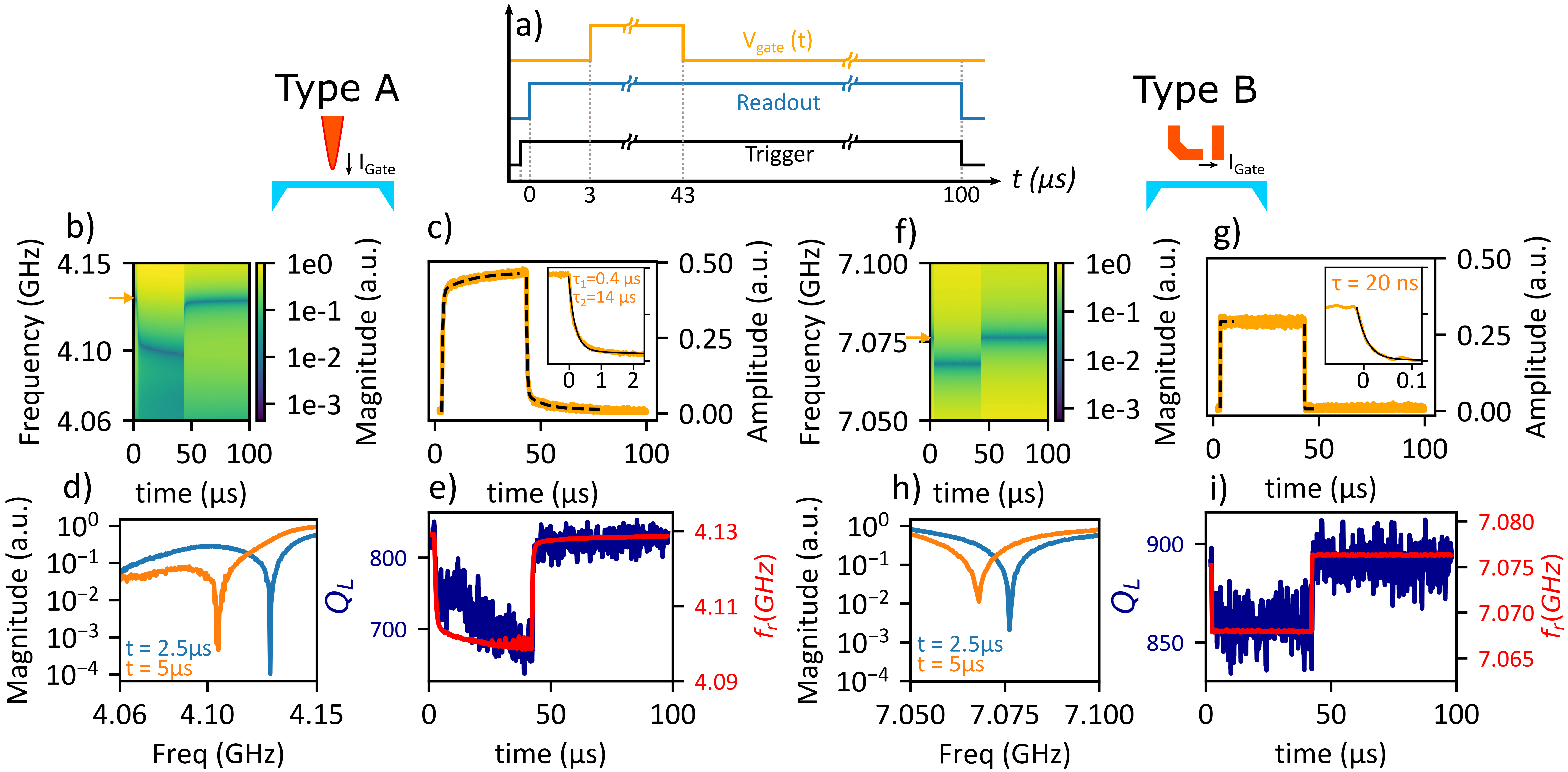}
\caption{\label{fig2} \textbf{Time Domain Characterisation Scattering Parameters}. \textbf{a} Pulse sequence of the measurement. Time trace showing the amplitude of $S_{21}$ for the type A device near the resonator frequency in \textbf{b} and type B in \textbf{f}. \textbf{c}, \textbf{g} Response of the resonator to the gate pulse, measured at the steady-state resonance frequency (indicated by the orange arrow), for types A and B, respectively. The insets zoom in on the fall time, where zero on the time axis corresponds to the end of the gate pulse. \textbf{d} and \textbf{h} line cuts from the dataset in \textbf{b} and \textbf{f} respectively, taken at two specific time stamps (before and after the gate actuation), illustrating the magnitude of the resonator's $S_{21}$. \textbf{e}, \textbf{i} Loaded quality factor, defined as $f_r/\delta f$ where $\delta f$ is the full width at half maximum of the $S_{21}$ magnitude dip and Time trace of the resonance frequency plotted along with the resonance frequency. The gate pulse amplitude used for \textbf{a}-\textbf{e} is 1.72V. }
\end{figure*}

\section{Time-domain Measurements}
After establishing the tunability of the resonances with $V_G$ using spectroscopic measurements, we now probe the devices in time domain to quantitatively evaluate the performance of both gate geometries. This step is crucial for assessing their potential in constructing devices suited for practical applications.

We perform the following sequence of measurements. At $t=0$, we apply a coherent tone of frequency $f_p$ at the waveguide for a duration of total $100~\mu\rm{s}$. At the same time, starting at $t=3~\mu\rm{s}$, we apply a voltage pulse to the gate of fixed duration of $40~\mu\rm{s}$ and varying amplitude [Fig.~\ref{fig2}(a)]. We continuously record the complex amplitude of the coherently scattered signal through the waveguide, with a sampling rate of 1~GS/s. (See Methods). We note that by recording time traces at different probe frequencies $f_p$, we obtain more information than by examining the response at a single frequency or tracking the resonant frequency in response to $V_{gate}$. This approach allows us to examine the transient aspect of the resonator's response to the perturbation caused by the gate pulse, providing a deeper understanding of the dynamic processes at play. In the limit in which the nanowire dynamics is slower than the time constant of the resonator, this time-resolved approach allows us to analyze both the resonator's time-dependent resonance frequency and its loaded quality factor, showing how the system responds and recovers from the applied gate pulse.

We first examine the time-dependent scattering parameters of the type A gate configuration, as shown in Fig.\ref{fig2}b-e. Specifically, line-cuts of $S_{21}$  taken at two different time stamps, one prior and the other following the gate actuation [Fig.\ref{fig2}d] reveal a decrease in $f_r$, consistent with our earlier spectroscopic measurements [Fig.~\ref{fig1}]. In our initial analysis, we focus on the time trace of the resonance frequency at $V_G=0$~V [Fig.\ref{fig2}c]. The time traces of $|S_{21}|$  near both the rise and fall edges is not described accurately by a single-exponential model. Instead, they necessitate a linear combination of two exponential functions, $C_1 e^{-t/\tau_1}+C_2 e^{-t/\tau_2}$, where $\tau_{1,2}$ represent distinct characteristic time-scales. For the initial rising part ($3\leq t \leq43~\mu\rm{s}$) of the trace measured at $V_{gate}\,=\,1.72\,V$, a double-exponential fit unveils a short time constant  ($\tau_{1, \rm{rise}}=450$~ns) followed by a longer one ($\tau_{2, \rm{rise}}$ = 14~$\mu\rm{s}$). This indicates the presence of at least two kinetically distinct processes affecting the resonator's response. 
Analysing line-cuts at nanosecond intervals yield real-time $f_r$ (determined from the minima of $|S_{21}|$ \textit{vs} $f_p$) and $Q_L$ (determined from the full-width at all maximum of $|S_{21}|$ at $f_p$). For $Q_{L}$ (Fig.\ref{fig2}e) we observe an initial rapid decay over 450~ns, followed by a prolonged more gradual decline up to the fall edge of the gate pulse (at $t=43~\mu\rm{s}$). Similarly, $f_r(t)$ demonstrates a comparable bi-exponential behavior with an initial fast decay of $\tau_1=450$~ns before transitioning to a slower one with characteristic time $\tau_2=14.5 \mu\rm{s}$. 

At the gate pulse's end, the decrease in $|S_{21}|$ also follows a dual-exponential decay, characterized by a short-time constant $\tau_{1, fall}=470$~ns and a subsequent lengthier one $\tau_{2, fall} =14.2\mu \rm{s}$. Additionally, $f_r$ follows a corresponding recovery pattern. For $Q_L$, a similar rapid initial response is observed; however, its recovery to the initial steady-state is prolonged, extending up to $34\mu \rm{s}$ after the gate pulse end, indicating that the kinetics governing the resonator's frequency shift ($L_K^{NW}$) operate more rapidly than those affecting its losses upon applying the gate voltage. For the type-B device (Fig.\ref{fig2}f-i), we observe a significantly faster resonator response to voltage pulses compared to type-A configuration. The time traces around both the rise and fall edges are best fit by single exponential functions, with time constants of $\tau_{rise}~=~27$~ns and $\tau_{fall}=26$~ ns [Fig \ref{fig2}g]. This rapid response is uniform across both $Q_L$ and $f_r$. For comparison, the resonant frequency and loaded quality factor of the type-B resonator give a characteristic time constant $Q_L/2\pi f_r = 20\approx\rm{ns}$. 

\begin{figure}
\includegraphics{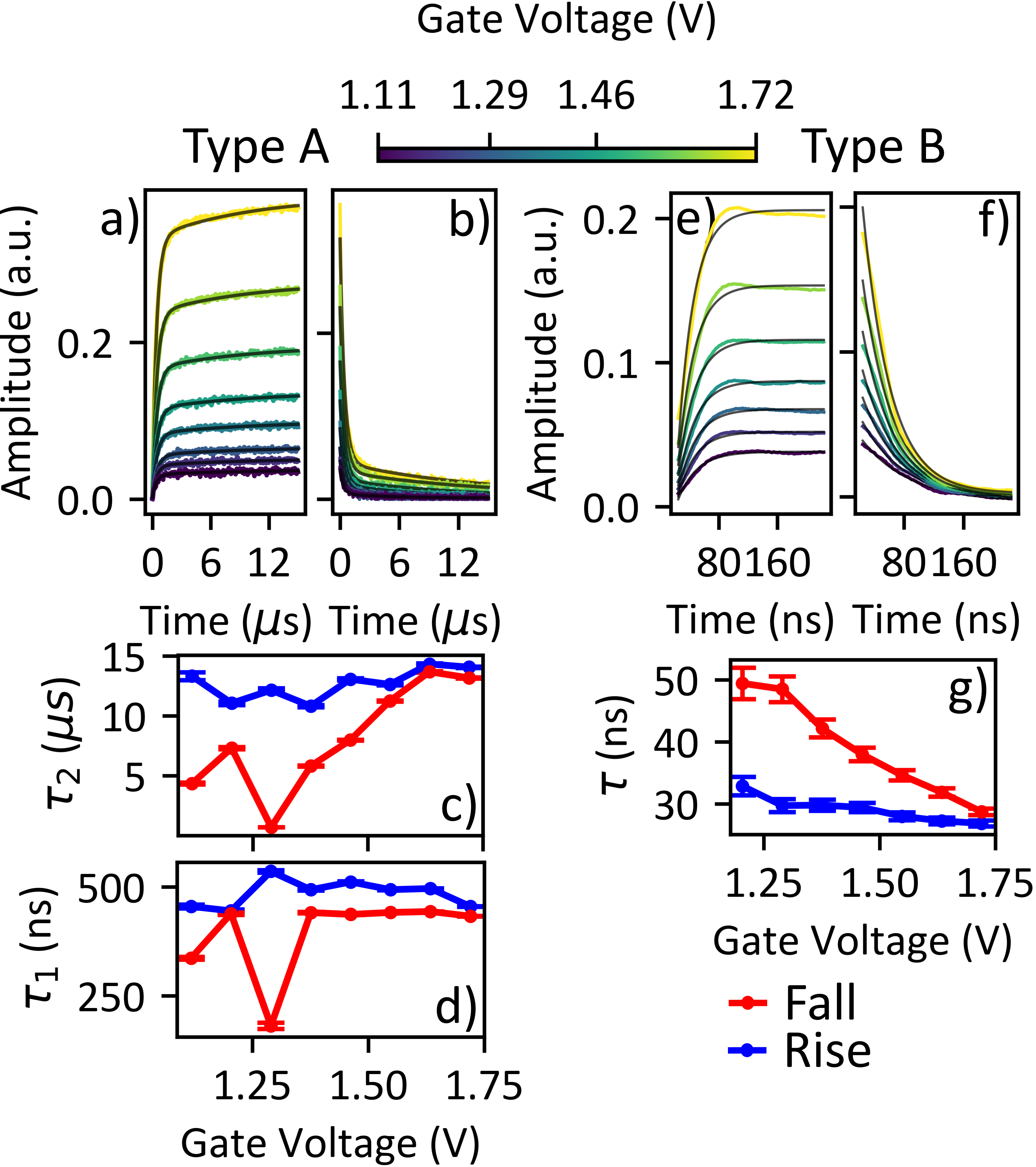}
\caption{\textbf{Response time of the resonator's frequency relative to gate pulse amplitude.} \textbf{a} and \textbf{b}, Time-domain traces and fits of the resonator response as a function of gate pulse amplitude for the type A device with \textbf{a} showing the onset of the gate pulse (rise) and \textbf{b} showing the fall. The zero on the time axis corresponds to the start or the end of the gate pulse, respectively. \textbf{c} and \textbf{d} Time traces for the type B device. \textbf{e}, \textbf{f}  Extracted rise (red) and fall (blue) time constants for the type A gate, with \textbf{e} showing the fast response time constants for both rise and fall and \textbf{a} showing the fast response time constants for both rise and fall and \textbf{b} detailing the slower response time constants for rise and fall. \textbf{g} Rise and fall time constants for the type B gate. \label{fig3}}
\end{figure}

Upon analysing the response to varying $V_G$ for the type-A device, the shorter response time, $\tau_1$ remains largely constant for both rise and fall edges, suggesting a stable initial response of the resonator to the gate pulse. In contrast, the longer response time, $\tau_2$, for the fall edge, monotonically increases with higher $V_G$ values, implying that the mechanisms governing the superconductivity relaxation in the nanowire decelerate as the voltage amplitude increases. A contrasting behavior is observed with the type-B gate configuration, where time constants for both the rise and fall edges decrease with an increasing $V_G$. Notably, the time constant for the fall edge shows a more pronounced reduction. This phenomenon can be attributed to the quasiparticle recombination rate being inversely proportional to the quasiparticle density, implying that higher quasiparticle densities lead to faster recombination rates \cite{wilson_time-resolved_2001}. It is important to note that the measurement of the tuning speed, based on changes in the scattering parameters, is constrained by the ring-up time of the resonator, quantified by $Q_{L}/\omega_0$. In our experiments, this duration is approximately 15 ns for the type A device and 25 ns for type B (Supplementary Fig.~S4) This constraint suggests that observed tuning dynamics, particularly for type B, are near the resonator's response limit, indicating potential for faster actuation not fully captured in our analysis.

\begin{figure}
\includegraphics{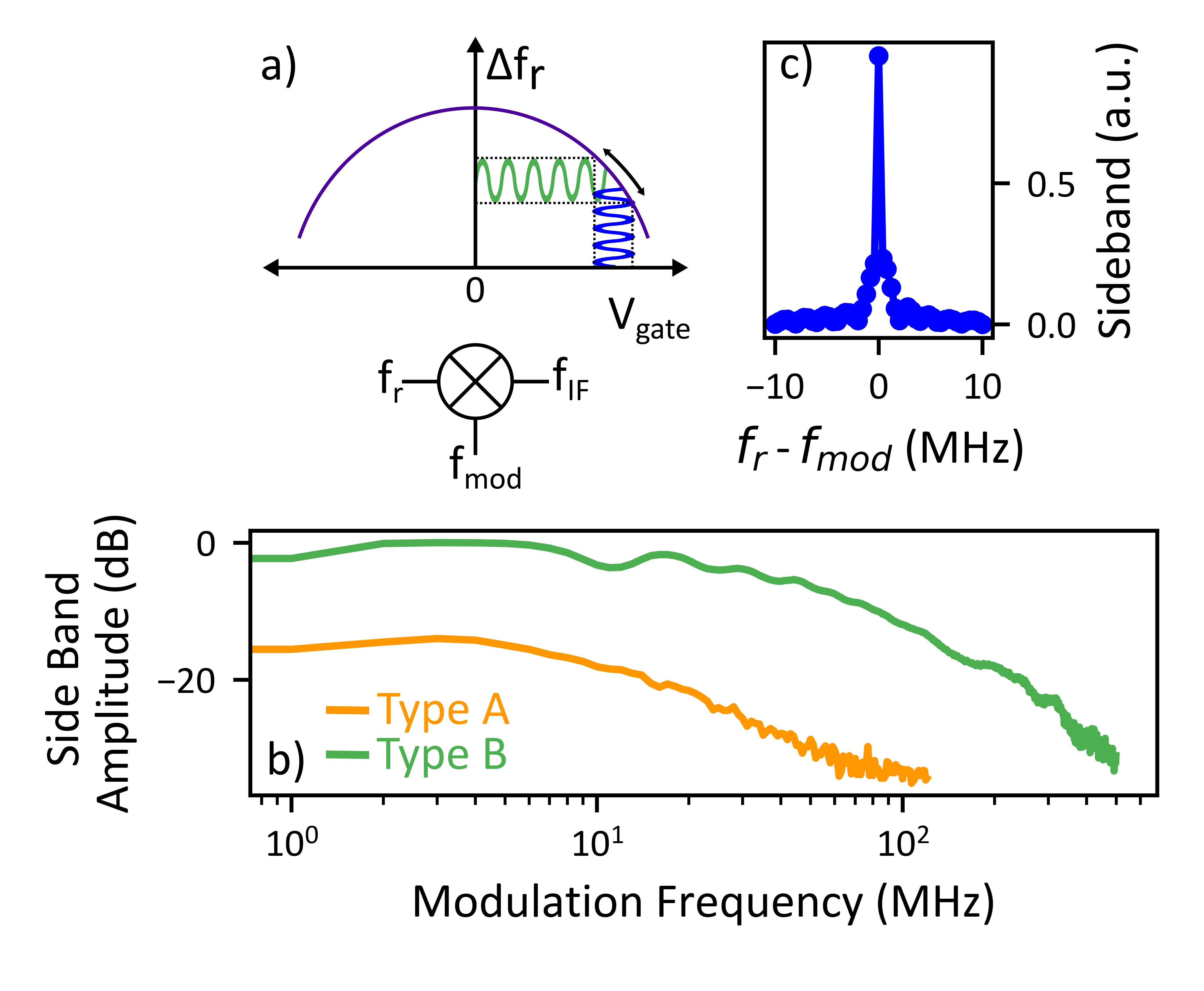}
\caption{Measurements of fast switching speeds conducted by mixing the signal from the resonator with the drive frequency, while a continuous wave (CW) tone and a constant DC bias are applied to the gate.  \textbf{a} Schematic diagram of the heterodyne measurement setup. \textbf{b} Sideband amplitude as a function of the modulation frequency for type A and type B gates. The DC bias for each gate type is optimized to achieve maximum modulation speed. \textbf{c} Sideband spectrum centered at the intermediate frequency (IF). \label{fig4}}
\end{figure}

To estimate how fast we can modulate the Dayem bridge in both gate configurations, we perform heterodyne measurements by mixing the signal leaking out from the modulated resonator at frequency $f_{mod}$ with the drive signal $f_r$ (Fig \ref{fig4}a). Put simply, the gate undergoes modulation at frequency $f_{mod}$, while we maintain a constant DC bias close to the onset gate threshold. The amplitude of the sidebands at the intermediate frequency ($f_{IF}\,=\,\lvert f_r-f_{mod} \rvert$, Fig.\ref{fig4}b), resulting from this mixing, is recorded and plotted as a function of $f_{mod}$ (Fig.\ref{fig4}c). The presence of sidebands in the spectra is indicative of a coherent process. In the type A gate configuration, modulation sidebands were observed up to 120 MHz, corresponding to a modulation period of 8.3 ns, with the -3dB cut-off at 12 MHz indicating the effective bandwidth of the system. As anticipated, the type B gate exhibited a more rapid modulation capability, with a -3dB cut-off at 37 MHz and modulation sidebands extending up to 500 MHz corresponding to a modulation period of 2~ns.
We believe that the observed cutoff for type B is limited by attenuation in the lossy coaxial cable used to drive the gate (see Suppl.~Mat.), suggesting that it could be possible to modulate the type-B device faster than reported.
We observe that the sideband amplitude for the type A gate is 18 dB weaker than that for the type B gate, indicating a more efficient tuning mechanism associated with type B.

\section{ Discussion and Conclusion}
In our study, we explored the dynamics properties of a Dayem bridge integrated within a microwave resonator, highlighting how gate geometry critically affects the speed at which an NW can be actuated with a voltage pulse. The traditional finger-type gate (type-A)\cite{desimoni2018, elalaily2021, ritter2021, koch_gate-controlled_nodate, puglia2021}, presents a
slower switching response, with a limitation around $\sim15\mu\rm{s}$, making it less suitable for high-speed device application. Conversely, employing a set of remote electrodes positioned $1\,\mu m$ away from the nanowire enhances the response speed, achieving modulation frequencies up to 500 MHz. This configuration stands out as the preferable configuration to develop advanced high-speed superconducting logic devices. The modulation of superconductivity with gate voltage in both types of devices is directly linked to the generation of leakage current \cite{elalaily2023d}. However, the flow and the impact of this leakage current differ based on the gate geometry. In type-A devices, leakage current directly flows to the nanowire, resulting in a direct injection of quasiparticles with energy ($\approx eV_{gate}$) significantly larger than the superconducting gap of Nb \cite{mccaughan2014, mccaughan2019b}. This leads to the formation of hot-spots, which then equilibrate through a mix of recombination and diffusion processes out of the nanowire \cite{kaplan_quasiparticle_1976, wilson_time-resolved_2001}. 

In contrast, the type-B gate's design, featuring an adjacent ground electrode and its overall distant placement, effectively suppresses the direct quasiparticle injection into the nanowire. Instead, a phonon flux, originating near the gate propagates through the substrate, leading to the breaking of Cooper pairs within the nanowire \cite{semenov_spectral_2005, semenov_quantum_2001}. The maximum energy carried by longitudinal acoustic phonons at the Brillouin zone boundary in Si is around 60 meV~\cite{giannozzi_ab_1991}, two orders of magnitude smaller than the energy of quasiparticles directly injected in the case of Type-A geometry. The cascading from the pair-breaking phonons' energy to sub-gap quasiparticles is faster in this case. The superconductivity restores as the quasiparticle density return to it's thermal equilibrium~\cite{patel_phonon-mediated_2017}.  

The observed time scales in modulating the resonator's resonance frequency reflect these disctint processes. The presence of two time scales, significantly differing in magnitude, suggests the coexistence of both processes in modulating the resonances \cite{elalaily2023d}. It is noteworthy that quasiparticle injection, whether direct injection (type-A) or phonon-mediated (type-B), increases with the gate voltage's magnitude. However, the measured time-scales diverge significantly between gate architecture: the type A device's lower fall time-scale $\tau_2$ increases with $V_G$ whereas the type-B device's fall time, $\tau$, decreases with $V_G$. Further experimental investigations could shed light on the complex interplay of mechanisms behind the voltage-induced superconductivity modulation. 

Our findings offer an insight into the complex temporal dynamics within such devices, potentially paving the way for innovative superconducting electronic application. The limited switching frequency of type-B devices, constrained by the thermo-coax's bandwidth used in our measurements, suggests room for improvement. By adopting on-chip filtering and measurement lines with broader bandwidths, we could potentially enhance the observable switching frequency threshold. Such rapid tunability of microwave resonances could lead to the development of on-chip superconducting microwave switches, offering minimized cross-talk, a common issue in flux-controlled devices. Our research lays the groundwork for advancing superconducting technologies through these innovative device architectures.

\section{Acknowledgements}

S.G. thanks A.~Fuhrer for useful discussions. K.R.A. and F.J. thank Axel M. Eriksson for help in setting up time-domain measurements. F.J. thanks Mikael Kervinen for useful discussions regarding the fabrication of the devices. F.J. also thanks Niclas Lindvall and Henrik Frederiksen for their support in the cleanroom during the fabrication of the devices. The chips were fabricated in the Chalmers Myfab cleanroom. This work was financially supported by the European Research Council via Grant No.~ 964398 SuperGate and Grant No.~101041744 ESQuAT, and by the Knut and Alice Wallenberg Foundation via the Wallenberg Centre for Quantum Technology (WACQT).
 
\appendix

\section{Fabrication}
The devices were fabricated at the Nanofabrication Laboratory in Chalmers using high-resistivity ($\rho > 10 k\Omega.cm$) n-type undoped (100) 2-inch silicon wafers. Before film deposition, the wafers underwent deoxidization in an HF bath. A 20 nm thick Nb film was then sputtered using a magnetron sputtering tool. The patterning of the resonators and feedline was achieved using a DWL 2000 laser writer from Heidelberg Instruments. Post-development, the Nb film was etched using a Plasma-lab 100 from Oxford Instruments, employing a reactive ion etching (RIE) system. The etching process utilized a $Cl_2/Ar$ gas flow at 40 sccm/5 sccm with an RF-field power of 50 W. Residual resist was subsequently removed using an ultrasonic bath with photoresist remover 1165, acetone, and IPA.

The Dayem bridges and gates were patterned using a 100-keV EBPG5200 electron beam lithography (EBL) system on a single layer of PMMA 950 A4 resist. The optimal dose determined was 880 $\mu \rm{C}/\rm{cm}^2$. Proximity effect correction (PEC) was employed to prevent shorting between the nanowire (NW) and the finger gate. Post-exposure, the development was carried out in MIBK:IPA (1:3) for 90 seconds. Any remaining resist residues were cleaned using photoresist remover and acetone in an ultrasonic bath. Finally, the 2-inch wafer was diced into individual chips measuring 7mm x 7mm. These chips were then packaged and connected with Al wires for testing and application.

\section{Measurement Set up}
A schematic of the measurement setup is shown in Supplementary Fig.~S1. Measurements are conducted in BlueFors BFLD250 dilution refrigerator. The sample is embedded in an 8-port sample holder and connected to the coaxial cables by aluminum wirebonds. Probe signals are sent from a vector network analyzer (port 1) attenuated by -63 dBm with attenuation at each plate as noted. The signal passes through a bandpass filter, then through the sample, and is returned through another filter. It then passes through an isolator with 20 dB isolation and 0.2 dB insertion loss, and is amplified by a low-noise amplifier mounted to the 4K plate, as well as by two room-temperature amplifiers (Pasternack PE1522). The gate electrode is connected to a voltage source and passed through a low-pass filter, mounted at the voltage source, and connected to the device with thermocoax cables.
Time-domain measurements are performed using a microwave transceiver, the Presto model from Intermodulation Product, with a sampling rate of 1 GS/s. To enhance the signal-to-noise ratio, we perform between $10^4$ and $10^6$ averages, depending on the drive power.

\nocite{*}

\bibliography{apssamp}% Produces the bibliography via BibTeX.

% If you had more sections, they would go here.

% No need to switch back to two-column grid if this is the end of your document.

\end{document}



\begin{figure*}
\subsection*{Supplementary 1}
\includegraphics[width=\textwidth]{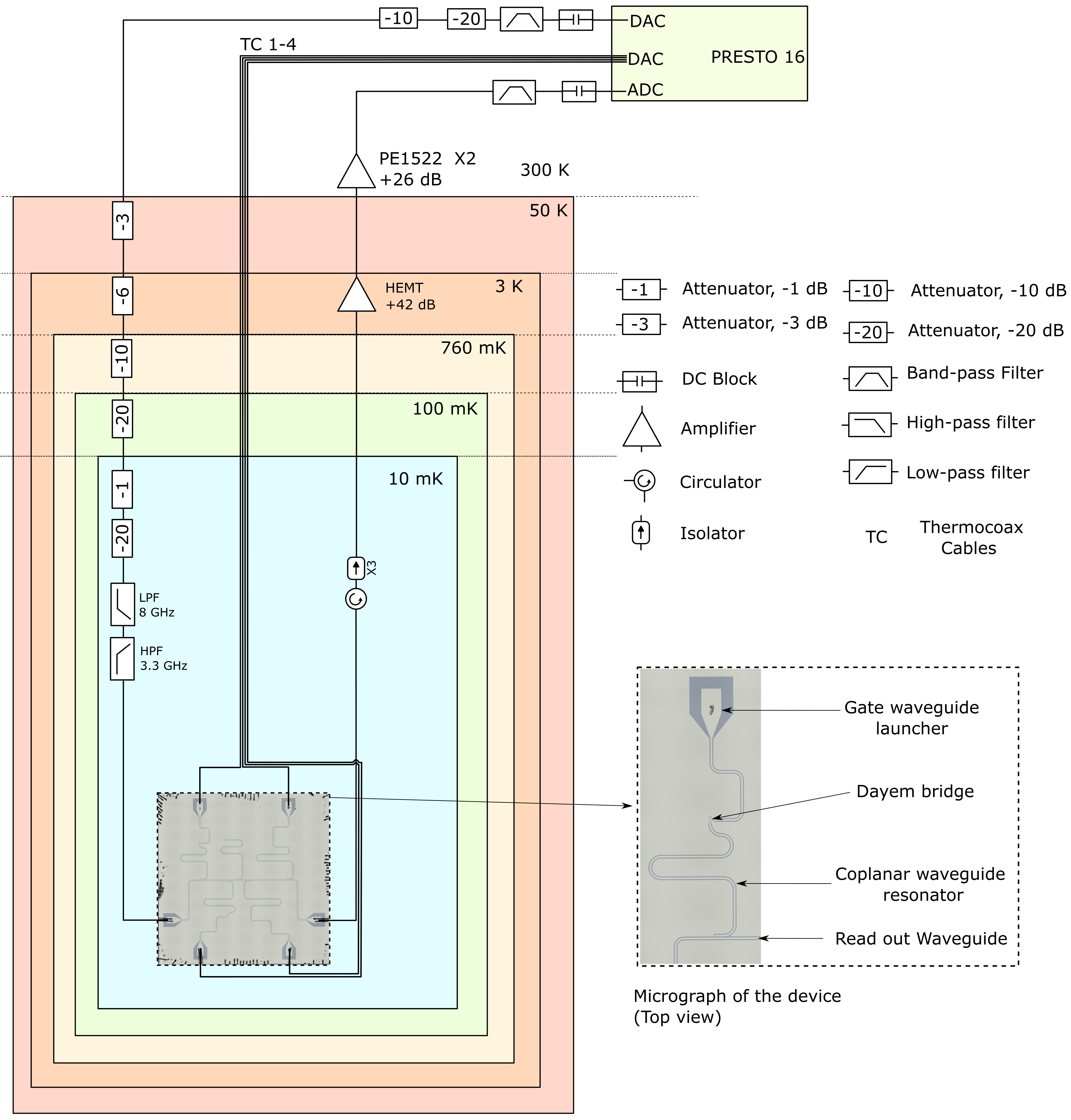}
\caption{\label{figS1}\textbf{Transmission measurement setup}. An AWG (Arbitrary Waveform Generator) Presto 16 is connected to both a highly attenuated input line and an amplified output line to probe the resonators. The gates on each resonator are linked to the AWG via thermocoax cables.}
\end{figure*}
\newpage
\subsection*{Supplementary 2}
\begin{figure*}
\includegraphics{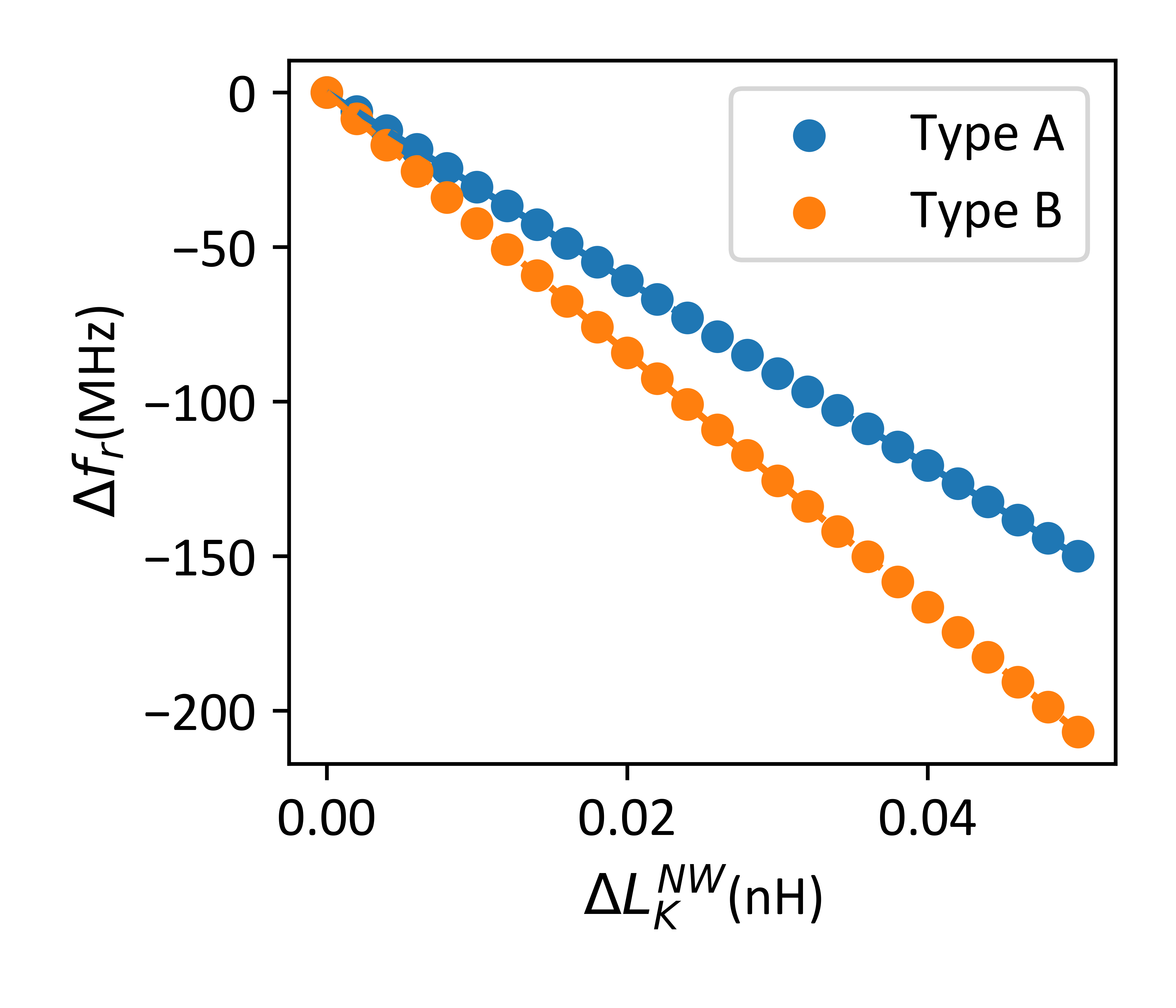}
\caption{Relative change in resonator frequency for types A and B as a function of the relative change in the nanowire's kinetic inductance.\label{figS2}}
\end{figure*}
\subsection*{Supplementary 3}
\begin{figure*}
\includegraphics{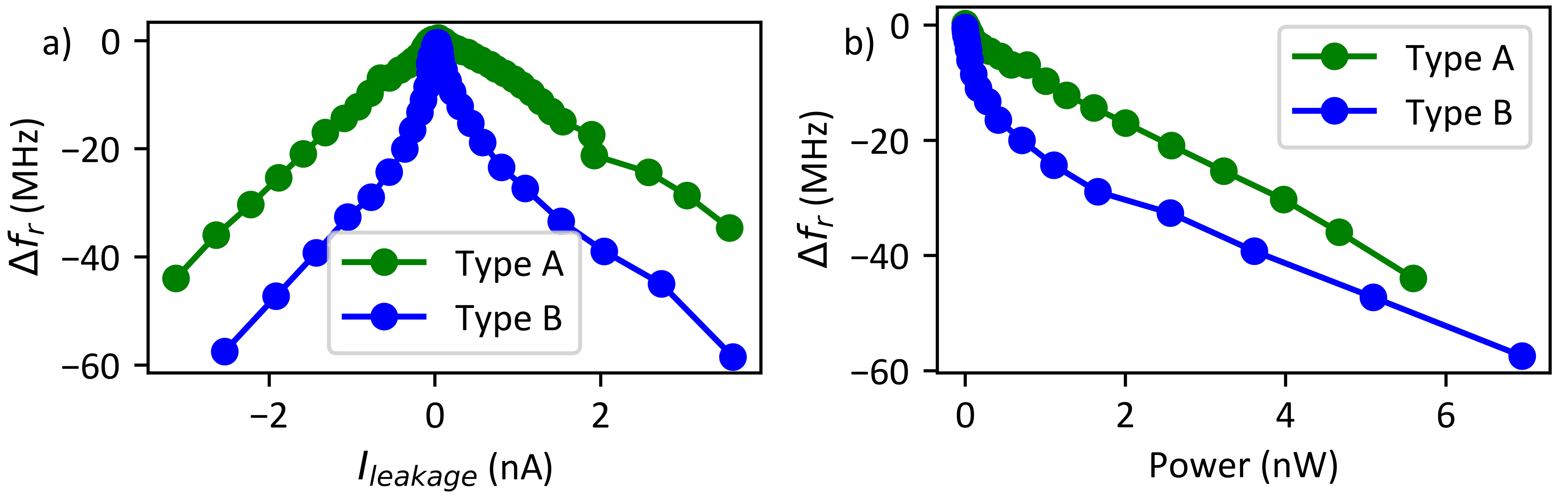}
\caption{a) Relative change in resonator frequency as a function of gate leakage current for types A and B. b) Comparison of Frequency Change as a function of dissipated power by type A and B gate. \label{figS3}}
\end{figure*}
\newpage
\subsection*{Supplementary 4}
\centering
\begin{figure*}
\includegraphics{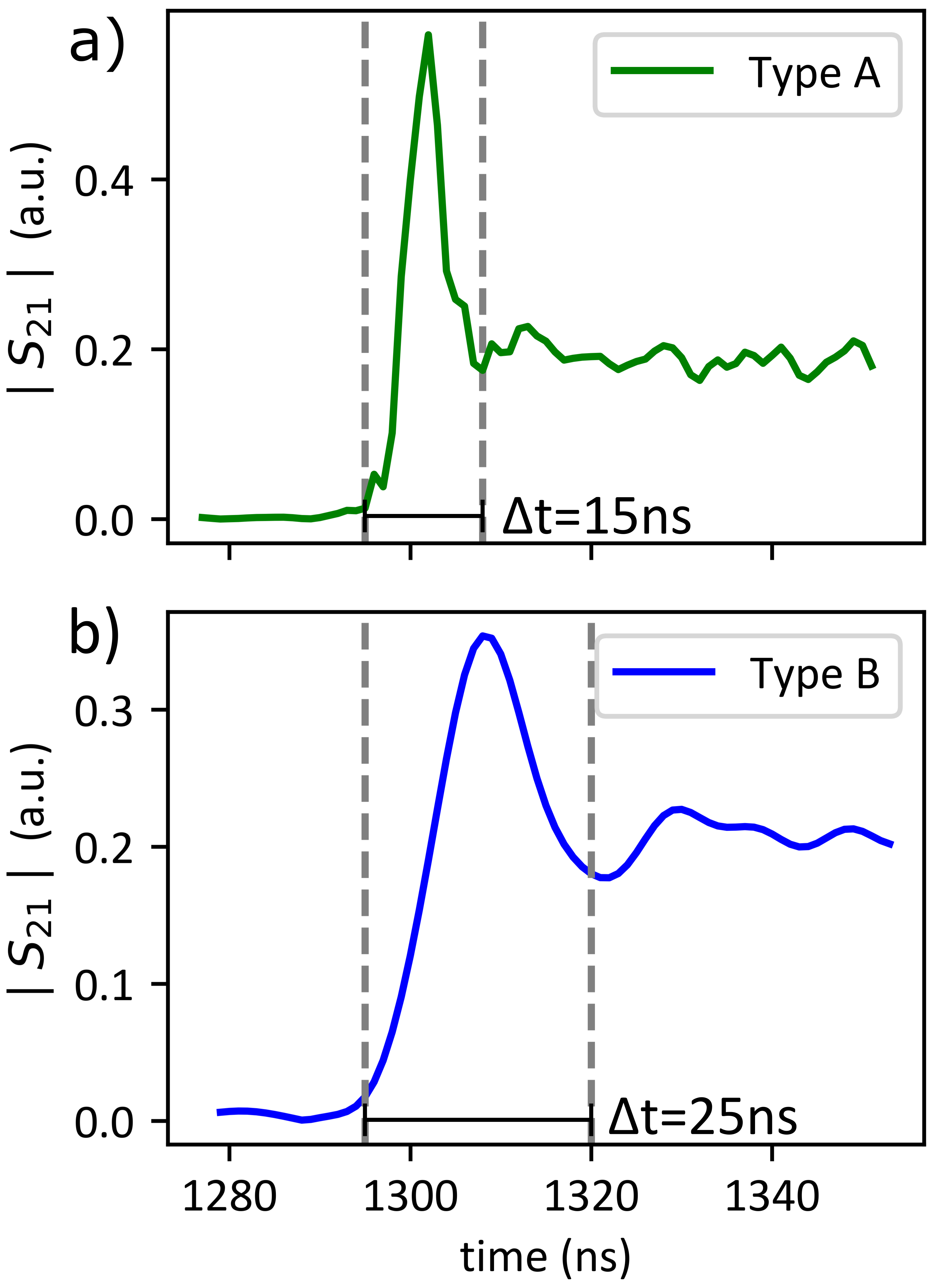}
\caption{Ring-up time of the resonator, illustrating the response speed of Type A (a) and Type B (b) resonators. \label{figS3}}
\end{figure*}
\cleardoublepage